\documentclass[journal]{IEEEtran}

\usepackage[T1]{fontenc}
\usepackage{graphicx}
\usepackage{cite}
\usepackage{subcaption}

\usepackage{overpic}
\usepackage{amssymb}
\usepackage{amsmath}
\usepackage{amsthm}
\usepackage{array}
\usepackage{color}
\usepackage{url}

\IEEEoverridecommandlockouts

\theoremstyle{plain}

\newtheorem{lemma}{Lemma}
\newtheorem{corollary}{Corollary}
\newtheorem{remark}{Remark}

\def\Ttran{\mbox{\tiny $\mathrm{T}$}}
\def\CN{\mathcal{N}_{\mathbb{C}}} 

\def\be{{\boldsymbol{e}}}

\newcommand{\pp}[1]{{\left( #1 \right)}}

\begin{document}

\title{Intelligent Reflecting Surfaces:\\ Physics, Propagation, and Pathloss Modeling}

\author{
\IEEEauthorblockN{{\"O}zgecan {\"O}zdogan, \emph{Student Member, IEEE}, Emil Bj{\"o}rnson, \emph{Senior Member, IEEE}, Erik G. Larsson, \emph{Fellow, IEEE}
\thanks{
\newline\indent The letter was supported by ELLIIT and the Swedish Research Council. 
\newline \indent The authors are with the Department of Electrical Engineering (ISY), Link\"{o}ping University, SE-58183 Link\"{o}ping, Sweden.
Email: \{ozgecan.ozdogan,emil.bjornson,erik.g.larsson\}@liu.se.}
}}

\maketitle

\begin{abstract}
Intelligent reflecting surfaces can improve the communication between
a source and a destination. The surface contains metamaterial that is
configured to ``reflect'' the incident wave from the source towards
the destination. Two incompatible pathloss models have been
used in prior work. In this letter, we derive the far-field pathloss
using physical optics techniques and explain why the surface consists
of many elements that individually act as diffuse scatterers but can
jointly beamform the signal in a desired direction with a certain
beamwidth. We disprove one of the previously conjectured pathloss
models.
\end{abstract}


\IEEEpeerreviewmaketitle

\vspace{-2mm}

\section{Introduction}

Conventional wireless communication systems consist of a transmitter
sending information-bearing electromagnetic waves to a receiver via an
uncontrollable propagation environment.  When searching for beyond 5G
network architectures, there is a growing interest in creating
real-time reconfigurable propagation environments
\cite{Wu2019a,Liaskos2018a,Bjornson2019d,Renzo2019a}. This can
potentially be achieved by  deploying special surfaces, known as 
\emph{intelligent reflecting surfaces (IRS)} \cite{Wu2019a},
\emph{software-controlled metasurfaces} \cite{Liaskos2018a,Bjornson2019d,Renzo2019a},   and \emph{reconfigurable
  intelligent surfaces} \cite{Basar2019a, Huang2019 }, that can control how the waves that reach them are reflected. In this context, the word
``reflection'' has a wide meaning \cite{Liang2015a}, including diffuse
reflection (e.g., scattering by rough material) and ideal specular
reflection (e.g., from an infinite mirror).

While the design of reconfigurable surfaces has a long history in the
electromagnetic literature \cite{Shaker2014a}, the communication
analysis is in its infancy. There is no consensus on the basic
propagation modeling, but two incompatible pathloss models have been
conjectured without derivation from physical principles: surfaces
consisting of many scattering elements \cite{Wu2018a, Bjornson2019IRS}
and surfaces consisting of many ideal mirrors \cite{Basar2019a}. In
this letter, we fill this gap by first explaining in
Section~\ref{eq:preliminaries} how a passive metallic surface scatters
an incident wave and then derive in Section~\ref{sec:IRS} how an IRS
must be designed to mimic such a surface while also controlling the
directivity of the scattered wave. This results in a rigorous pathloss model and a way to build system models that can be used for further research.

\vspace{-2mm}

 \section{Preliminaries: Passive Metallic Surface}
 \label{eq:preliminaries}

In this section, we review preliminaries on the field strength and
beamwidth of the waveform scattered by a ``passive'' metallic
(perfectly conducting) plate of finite size. The results will later be
used to explain the ideal operation of an IRS.

\begin{figure}[t!]
	\centering \vspace{-4mm}
	\begin{overpic}[width=.8\columnwidth,tics=10]{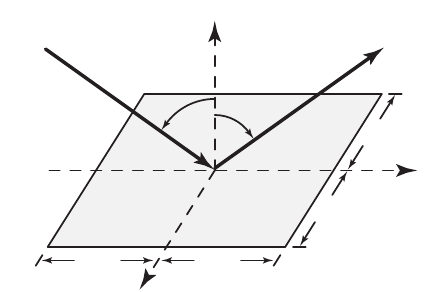}
	\put(19,5.5){$b/2$}
	\put(47,5.5){$b/2$}
	\put(84,35){$a/2$}
	\put(72.5,18){$a/2$}
	\put(42,37){$\theta_i$}
	\put(54,41){$\theta_s$}
	\put(35,-1){$\be_x$}
	\put(95,30){$\be_y$}
	\put(51,62){$\be_z$}
	\put(0,58){Incident wave}
	\put(72,58){Scattered wave}
\end{overpic} 
	\caption{An incident wave is scattered by a $a \times b$ metal plate.}
	\label{figurePlate}  \vspace{-5mm}
\end{figure}
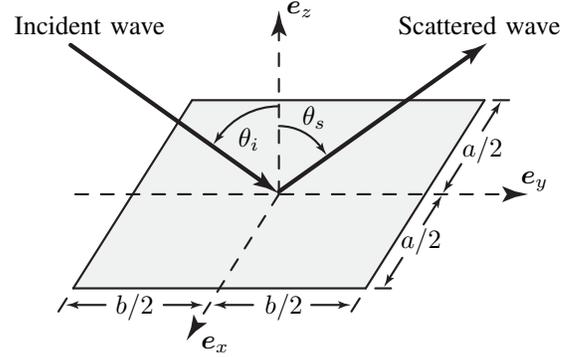

We consider a rectangular, perfectly conducting plate of size $a
\times b$, and negligible thickness, located in the horizontal plane
(spanned by $\be_x, \be_y$).  A point source far away, at distance
$d_i$, is radiating a linearly polarized electromagnetic wave with
wave number $ k=\frac{2\pi}{\lambda}$ where $\lambda$ is the
wavelength.  We assume, for the sake of argument, that the
polarization of the source is such that the E-field is parallel to
$\be_x$ and the H-field lies in the plane spanned by $\be_y$ and
$\be_z$.  Let $\theta_i \in [0,\frac{\pi}{2}]$ denote the angle of
incidence, that is, the angle between the Poynting vector of the wave
and $\be_z$. This setup is illustrated in Fig.~\ref{figurePlate}.
  
We further assume that $d_i$ is sufficiently large relative to $a$ and $b$  (i.e., the source is in the far-field) that the curvature of the wavefront, over the dimensions of the
plate, can be neglected. Hence, the impinging wavefield is
approximated as a plane wave with some magnitude $E_i$.  To determine when this
approximation holds, we compute the phase differential between a
spherical wave and its plane-wave approximation between the center and
the edges of the plate.  For the sake of argument, suppose the plate
is oriented such that the Poynting vector enters with $\theta_i = 0$
at the plate center; however the analysis can be generalized to
arbitrary incidence angles.  With the plane-wave approximation, the
wave from the source travels a distance of $d_i$ to any point on the
plate, but the wave is actually spherical and at the edges of the
plate it has traveled the distance $\sqrt{ d_i^2 + \frac{b^2}{4} }$.
The phase discrepancy is
\begin{align} 
k\pp{\sqrt{ d_i^2 +  \frac{b^2}{4}   } - d_i } \approx \frac{\pi}{4} \frac{b^2}{\lambda d_i} 
\end{align} 
With e.g., $b=1$\,m, $d_i=100$\,m, $\lambda=0.1$\,m, this is less than 
five degrees and should have a minor
impact. The incident plane wave has the electric and magnetic field distributions 
\begin{align}\label{incident:field}
&\!\!\!\mathbf{E}_i = E_i e^{-j k (\sin (\theta_i) y- \cos (\theta_i) z) } \be_x, \\
&\!\!\!\mathbf{H}_i \!=\! -\frac{E_i}{\eta}\left(  \cos(\theta_i)  \be_y + \sin(\theta_i) \be_z \right) e^{-j k (\sin (\theta_i) y- \cos (\theta_i) z) } ,
\end{align}
where $\eta$ is the characteristic impedance of the medium.

The E-field will induce motion of electrons in the plate. The
electrons will move in the direction of $\be_x$, but not in $\be_y$
(since the E-field is orthogonal to $\be_y$) and also not in the
$\be_z$-direction since the plate is assumed thin.  The moving
electrons induce electromagnetic radiation, resulting in a scattered
wave.

  \begin{lemma} \label{lemma:balanis}
 The squared magnitude of the scattered field, in the $\be_y,\be_z$ plane and at an arbitrary observation angle $\theta_s \in [0,\frac{\pi}{2}]$ (measured against $\be_z$)  is 
 \begin{align} \label{eq:field-strength}
S(r, \theta_s) \! = \!   \pp{\frac{ab}{\lambda}}^{\!\!2} \frac{E_i^2}{r^2}  \cos^2(\theta_i)
 \pp{ \! \frac{\sin\pp{ \frac{\pi b}{\lambda} (\sin(\theta_s) \!-\!\sin(\theta_i)) }}{ \frac{\pi b}{\lambda}(\sin(\theta_s) \!-\!\sin(\theta_i))} \! }^{\!2}  
 \end{align} 
 at a far-field observation distance $r \geq \frac{2 \max( a^2,b^2 )}{\lambda}$.
 \end{lemma}
 \begin{IEEEproof}
 This result follows from standard physical optics techniques (which neglect edge effects) as in \cite[Example 11-3]{Balanis2012a}.
 \end{IEEEproof}
 
The expression in \eqref{eq:field-strength} for the magnitude of the
scattered field in the far-field has several intuitive properties; for example, it is
proportional to the area $(ab)^2$ of the plate and to $E^2_i \propto
1/d_i^2$, since the impinging wave from a point source in
line-of-sight (LoS) has a field strength inversely proportional to
$d_i^2$. As expected from Snell's law, for the polarization that we
consider, $S(r, \theta_s)$ attains its maximum for the observation
angle $\theta_s = \theta_i$, which is the specular direction, for
which the last parenthesis in \eqref{eq:field-strength} is
unity.\footnote{For other incident wave polarizations, this is true
  only approximately.  However, the observation angle that maximizes
  $S(r, \theta_s)$ approaches $\theta_i$ as the size of the plate
  increases.}

\vspace{-2mm}
\subsection{Beamwidth of the Scattered Wave}
 
The expression $S(r, \theta_s)$ reveals that the scattered field looks
like a beam that tapers off as $\theta_s$ is moved away from
$\theta_i$.  The 3-dB beamwidth equals twice the
deviation $ |\theta_s-\theta_i|$ required to make the square of the
second parenthesis in \eqref{eq:field-strength} equal to $1/2$.
Using the Taylor expansion $\cos(x) = 1 + O(x^2)$ and standard
trigonometric identities,  we have
 \begin{align} \notag
&\sin(\theta_s)  - \sin(\theta_i)  = \sin(\theta_i + \theta_s - \theta_i) - \sin(\theta_i) \\ \notag
& =\sin(\theta_i) \cos(\theta_s-\theta_i) + \cos(\theta_i)\sin(\theta_s-\theta_i) - \sin(\theta_i) \\
& = \cos(\theta_i) (\theta_s-\theta_i) + O\pp{ (\theta_s-\theta_i)^2 } .  \label{eq:erik1}
 \end{align}
From \eqref{eq:erik1} and using the Taylor expansions $\frac{\sin(x)}{x} = 1 - \frac{x^2}{6} + O(x^3)$ and $( \frac{\sin(x)}{x} )^2 = 1 - \frac{x^2}{3} + O(x^3)$, we obtain the following second-order approximation of \eqref{eq:field-strength}:
 \begin{align} \notag
\!\! S(r,\theta_s) &\!= \!\pp{\frac{ab}{\lambda}}^{\!\!2} \frac{E_i^2}{r^2}   \cos^2(\theta_i) \pp{ \! 1- \frac{ \pi^2 b^2}{\lambda^2} \frac{\cos^2(\theta_i) (\theta_s-\theta_i)^2}{3} \!} \\
 &\quad+ O\pp{ (\theta_s-\theta_i)^3 }.
 \end{align}
 Hence, the 3-dB beamwidth is approximately twice the deviation $ |\theta_s-\theta_i|$
required to make $\frac{ \pi^2 b^2}{\lambda^2} \frac{\cos^2(\theta_i) (\theta_s-\theta_i)^2}{3} = \frac{1}{2}$:\vspace{-1mm}
 \begin{align}\label{eq:BW-3dB}
 |\theta_s-\theta_i| < \sqrt{\frac{1}{2} \frac{3 \lambda^2}{\pi^2b^2\cos^2(\theta_i)}} = \sqrt{\frac{3}{2}} \frac{\lambda}{\pi b\cos(\theta_i)}.
 \end{align}
This inequality shows that the 3-dB beamwidth is inversely
proportional to the plate width $b$. The beamwidth is also
proportional to the wavelength $\lambda$, thus a fixed-size plate can
provide an extremely narrow beamwidth in the visible spectrum (i.e.,
an almost perfect specular reflection), but 4-5 orders-of-magnitude wider
beamwidths in the typical radio spectrum bands.\footnote{Comparing radio signals
  at 6 GHz with green visible light at 600 THz, the former has a
  $10^5$ larger wavelength and thus needs a $10^5$ wider plate to give
  a reflection with the same beamwidth as for green light.}
  
  \begin{figure}[t!]
	\centering  \vspace{-2mm}
	\begin{overpic}[width=\columnwidth,tics=10,height=2.4in]{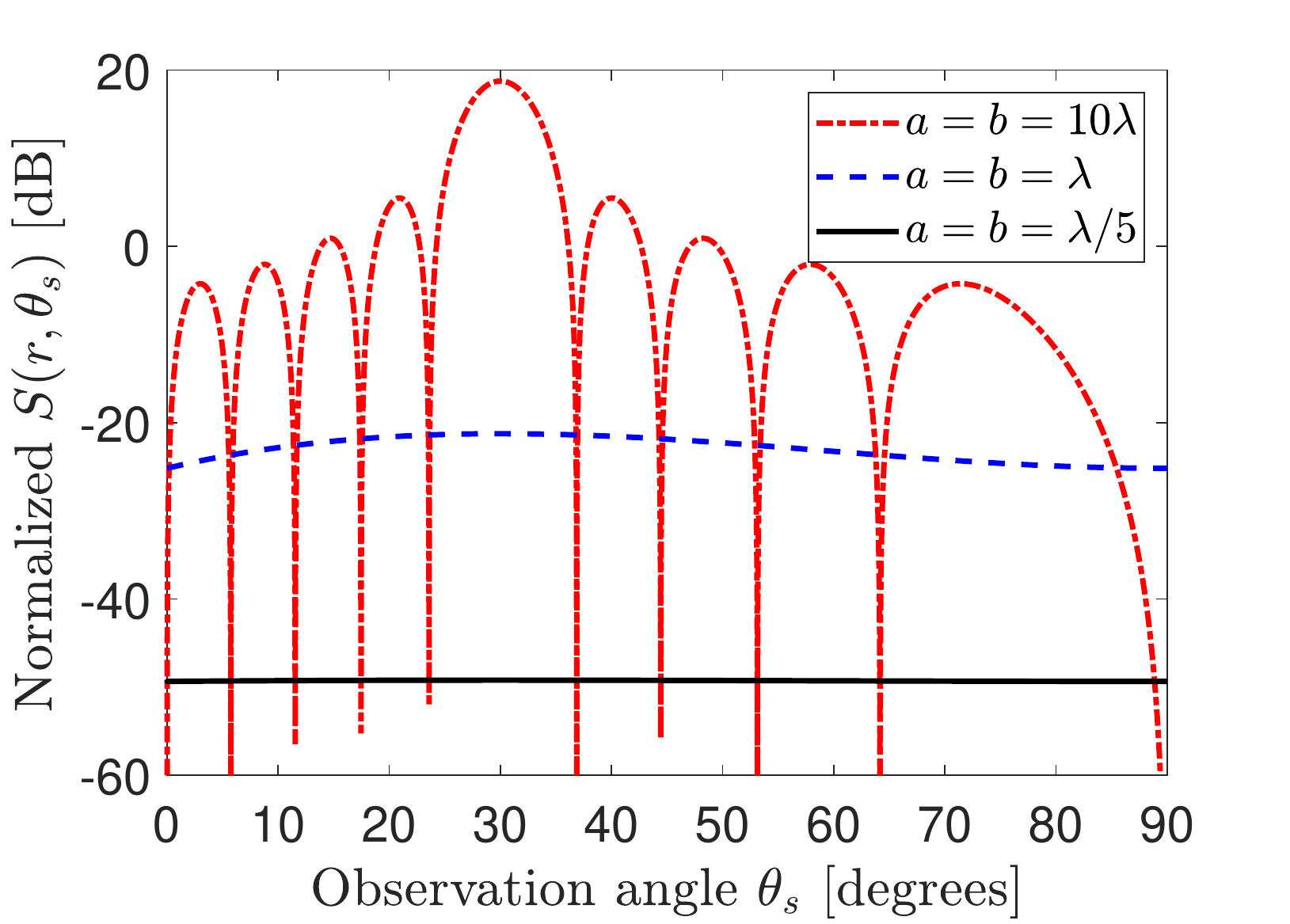}
\end{overpic} 
	\caption{Normalized squared magnitude of the scattered field
          as a function of the angle $\theta_s$ where  $\theta_i=30^\circ$.}\vspace{-4mm}
	\label{figureBeamWidth}
\end{figure}

The plate acts as a mirror in the sense that the diffusely reflected
or, more accurately, the scattered field has a 3-dB beamwidth
inversely proportional to $b$. Hence, it becomes very small if $b$ is
large relative to $\lambda$.  This scaling is no surprise: if we think
of the plate as an antenna, it is just the usual relation that
beamwidth is inversely proportional to antenna aperture.
  
Fig.~\ref{figureBeamWidth} shows the squared magnitude $S(r,
\theta_s)$ as a function of $\theta_s$. It is normalized by
multiplying with $\frac{r^2}{E_i^2}$, and we consider 
$\theta_i=30^\circ$. When $a$ and $b$ are smaller or equal
to the wavelength, the scattered field is almost equally strong in all
observation angles. It is first when the plate is substantially larger
than $\lambda$ that the beamwidth becomes small. The 3-dB beamwidth is
$\approx5.7^\circ$ for a plate with $a=b=10\lambda$ where the
second-order approximation in \eqref{eq:BW-3dB} gives
$\approx5.2^\circ$.
  
\begin{corollary}\label{cor1}
A receiving antenna of effective electrical size $\frac{\lambda}{v}
\times \frac{\lambda}{v}$, located at distance $r\gg b/v$ from the
plate with angle $\theta_s$ to the antenna center will receive the
signal power
 \begin{equation} \label{eq:received-power}
 S(r,\theta_s) \left( \frac{\lambda}{v} \right)^2.
 \end{equation}
\end{corollary}
\begin{IEEEproof}
The antenna will see the plate through an angular window of  $\frac{\lambda}{vr}$ radians.
As long as  
$\frac{\lambda}{vr} \ll \frac{\lambda}{b\cos(\theta_i)}$,
i.e., approximately, $r  \gg b/v$,
 the field strength will be approximately constant over the antenna and \eqref{eq:received-power} is obtained.
\end{IEEEproof}

Since $E^2_i \propto 1/d_i^2$ in LoS, Corollary~\ref{cor1} proves that
the received power is proportional to $(ab)^2/(d_i r)^2$, where the
proportionality constant depends on the wavelength and angles.  If $\theta_s = \theta_i$, the received power increases
monotonically with $a$ and $b$, because more energy is induced into
the plate and then radiated in a gradually narrower beam.  Even when
the beamwidth is small in absolute terms, it can be large relative to
$\lambda/r$ and, thus, most of the energy of the scattered field
anyway misses the receiver antenna aperture. This is why the
received power in \eqref{eq:received-power} is proportional to
$1/r^2$.  As $a$ and $b$ are increased without bound, eventually the
plane-wave approximation of the incident field breaks down and the
results in this section become inaccurate. In the asymptotic limit,
geometric optics can be used instead to model a perfect mirror.

\vspace{-2mm}
\subsection{Multiple Adjacent Metallic Surfaces}
\label{subsec:multiple-plates}

Since the plate has a finite size, we can deploy multiple adjacent
plates. If the gaps in between are sufficiently large then coupling
effects can be neglected and superposition applies.  When the
scattered fields from these plates are received at a given location,
the relative phase shifts will lead to constructive or destructive
interference.  Under ideal constructive interference, the squared
field strength from $N$ plates will be
\begin{equation} \label{eq:combination-field-strengths}
\left( N \sqrt{ S(r,\theta_s)} \right)^2 = N^2 S(r,\theta_s).
\end{equation}
The variables $N$, $a$, and $b$ only appear in
\eqref{eq:combination-field-strengths} as a joint term $(Nab)^2$,
where $Nab$ is the total area of the $N$ plates. Hence, it does not
matter if the total area is made up by many small or a few large
plates, the maximum received power is the same.
\vspace{-2mm}

\section{System Model for Intelligent Metasurfaces}
\label{sec:IRS}

If the incident angle $\theta_i$ from the transmitter to the surface
equals the observation angle $\theta_s$ leading to the receiver, then
the passive surface considered in Section~\ref{eq:preliminaries}
represents the ideal case. Since the angles are generally different
when one of the devices is mobile, the main purpose of an IRS is to
achieve ``anomalous reflection'' \cite{Liang2015a}, which means
shaping the scattered field so that the main beam is directed towards
the receiver.

We now consider an IRS consisting of a metasurface of the same
dimensions as in Fig.~\ref{figurePlate} and the same impinging plane
wave.  The goal of an IRS is to achieve total reflection with its main
beam pointing in a desired direction that we denote as
$\theta_r$. Hence, the surface must be designed to redirect the
incident plane wave $(\mathbf{E}_i, \mathbf{H}_i)$ in
\eqref{incident:field} and obtain the following ideal field
distributions of the reflected/scattered wave:
\begin{align}
&\!\!\!\!\!\mathbf{E}_r = E_r e^{-j k (\sin (\theta_r) y + \cos (\theta_r) z) } \be_x, \\
&\!\!\!\!\!\mathbf{H}_r \!\!=\!  -\frac{E_r}{\eta}\left(  \sin(\theta_r) \be_z \!- \cos(\theta_r) \be_y \right) e^{\!\!-j k (\sin (\theta_r) y + \cos (\theta_r) z) } \!.
\end{align}

A widely used approach in the literature that designs reflective
metasurfaces is based on the \emph{generalized Snell's law of reflection} \cite{Yu2011}.  Using this method, the required surface
phase profile to transform the incident wave $(\mathbf{E}_i,
\mathbf{H}_i)$ into $\left( \mathbf{E}_r, \mathbf{H}_r\right)$ is
obtained by tailoring the surface impedance.  At the surface ($z=0$),
the superposition of the incident and reflected E-field can be written
as \cite{Tretyakov2016}
 \begin{equation}
 \mathbf{E}_t = {E}_i  e^{-j k \sin (\theta_i) y  } \be_x + {E}_r  e^{-j k \sin (\theta_r) y  } \be_x.
 \end{equation}
Then, the desired phase of the desired reflection coefficient is
 \begin{equation}\label{phase:2}
 \phi_r(y) = \angle \left( \frac{{E}_r  e^{-j k \sin (\theta_r) y  }}{{E}_i  e^{-j k \sin (\theta_i) y  } }\right) = -k \sin (\theta_r) y +  k \sin (\theta_i) y ,
 \end{equation}
and differentiating it with respect to $y$ gives the gradient of the
reflection coefficient in the generalized Snell's law:
 \begin{equation}\label{phase:3}
 k (\sin (\theta_i) - \sin (\theta_r)) = \frac{d \phi_r (y)}{ d y},
 \end{equation}
which gives the relation between $\theta_i$, $\theta_r$ and the local
surface phase $\phi_r(y)$. By altering the surface impedance,
$\phi_r(y)$ is obtained at each point of the surface and the output
wave's desired phase $ - k \sin (\theta_i) y + \phi_r(y) = -k \sin
(\theta_r) y $ is achieved.

 \begin{figure}[t!]
 	\centering  \vspace{-2mm}
 	\begin{overpic}[width=1.05\columnwidth,tics=10,height=2.4in]{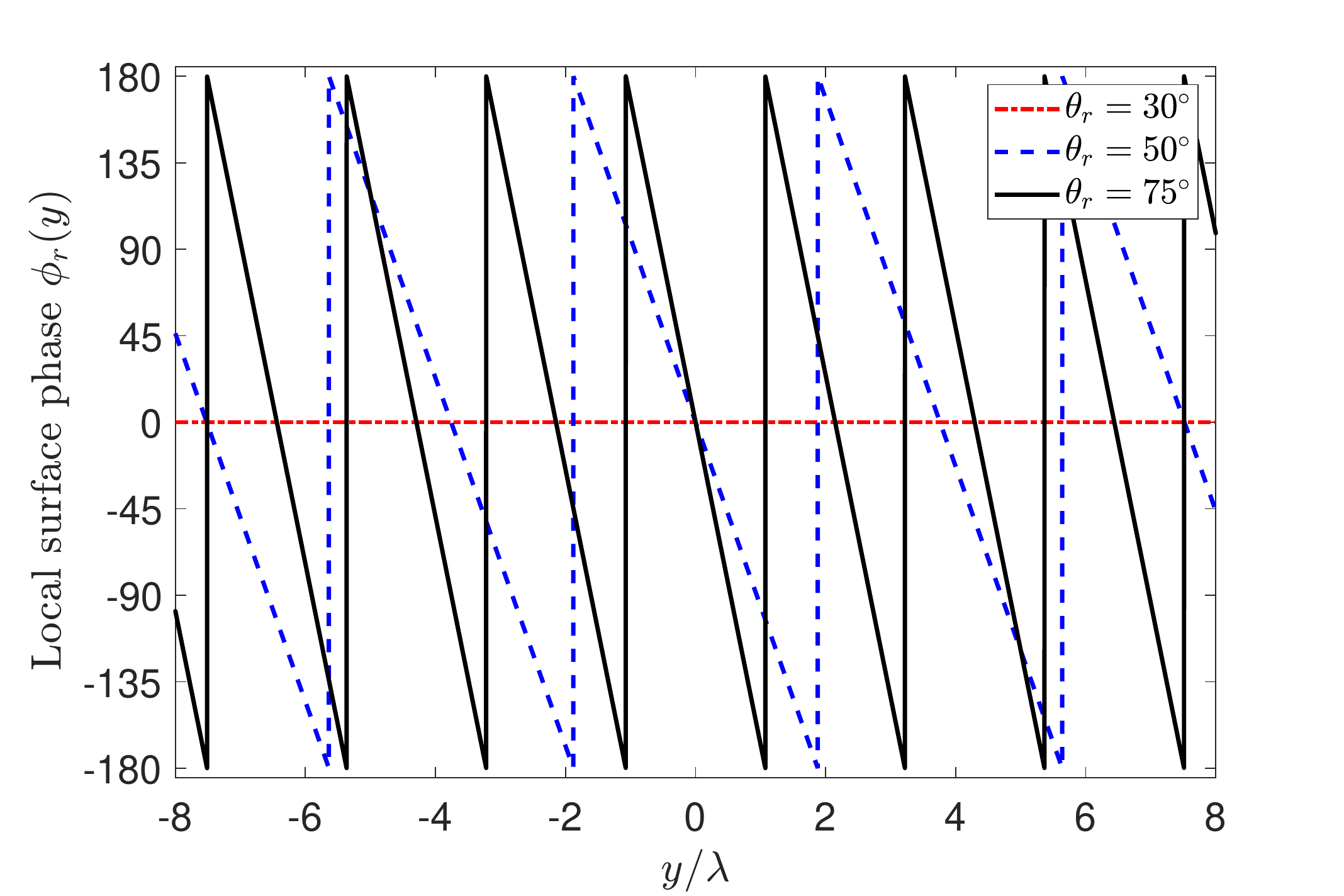}
 	\end{overpic} 
 	\caption{Local surface phase that is required to redirect the incident wave with $\theta_i=30^\circ$ in a desired direction $\theta_r$.}
 	\label{figureLocalPhase}\vspace{-5mm}
 \end{figure}
 
\begin{figure}[t!]
	\centering  \vspace{-4mm}
	\begin{overpic}[width=1.05\columnwidth,tics=10,height=2.35in]{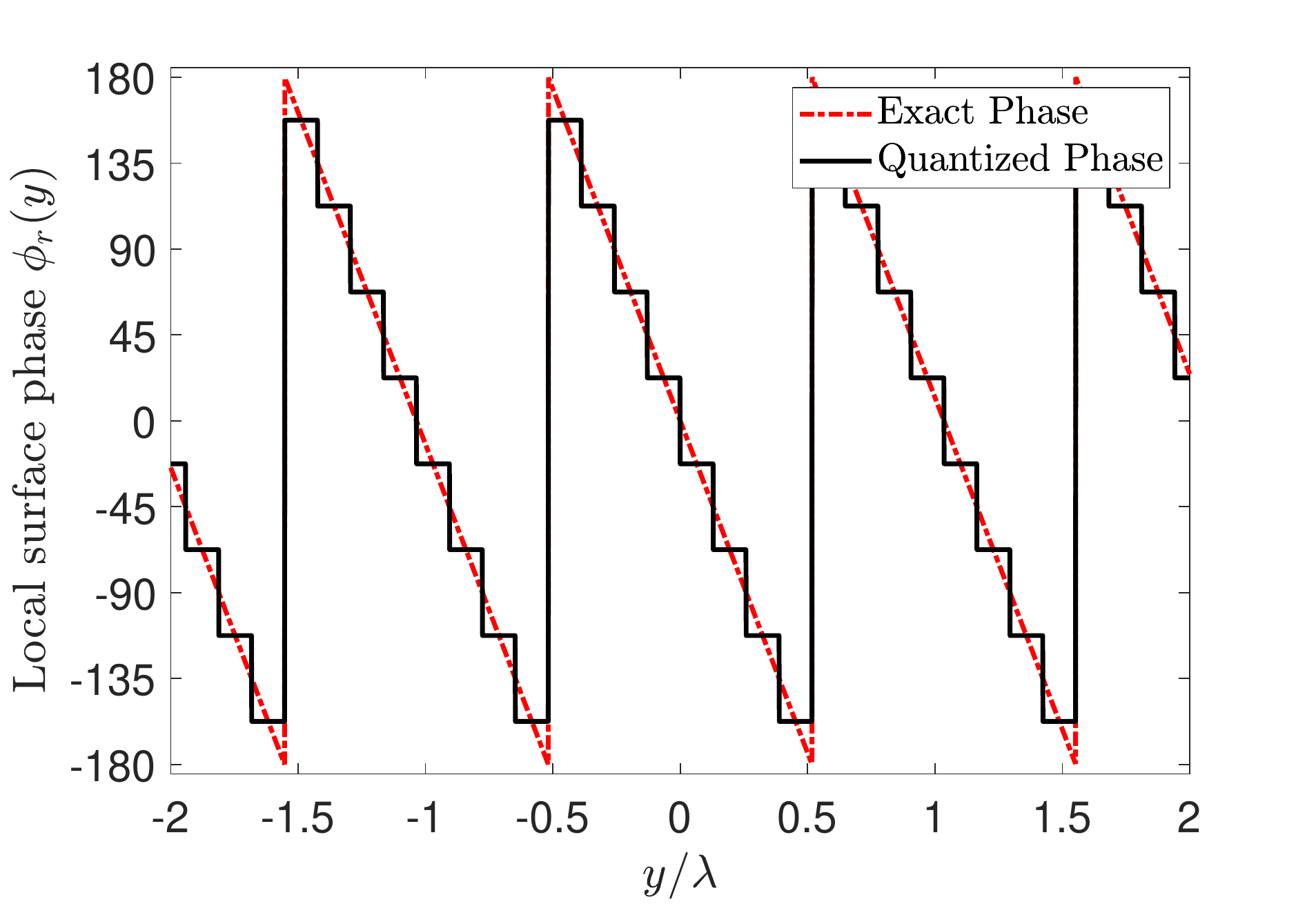}
	\end{overpic} 
	\caption{The quantized local surface phase that is required to redirect the incident wave with $\theta_i=0^\circ$ to $\theta_r = 75^\circ$. }
	\label{figureLocalPhaseQuantized} \vspace{-6mm}
\end{figure}

Fig.~\ref{figureLocalPhase} shows the required local surface phase
profile to redirect $(\mathbf{E}_i ,\mathbf{H}_i)$ with
$\theta_i=30^\circ$ into $(\mathbf{E}_r ,\mathbf{H}_r)$ for different
values of $\theta_r$.  
Fig.~\ref{figureLocalPhase} shows that it is easier to implement a surface when
the desired $\theta_r$ is close to $\theta_i$ since the required phase
profile varies more slowly over the surface (i.e., \eqref{phase:3} is
closer to zero). In practical implementations, the desired local phase
shift in the metasurface is discretized by dividing the surface into
sub-$\lambda$-sized elements, each having a constant phase shift. The smaller the
elements are, the more closely the local phase shift can be
approximated. In \cite{Headland2017}, the phase
distribution is discretized/quantized with a step size of $\lambda/5$,
to represent the finite number of resonators used in practice. In
\cite{Alu2016}, the surface is discretized by $\lambda/8$ as also
shown in Fig.~\ref{figureLocalPhaseQuantized}. 

Small high-precision elements require sophisticated design and expensive hardware, and also lead to coupling issues. On the other hand, if the elements are too large (relative to $\lambda$) then the required local phase will be coarsely quantized.  This will lead to a mismatch between the desired reflection angle and the surface response. In this paper, we neglect the errors due to quantization.

\vspace{-4mm}
\subsection{Propagation and Pathloss Model}

In contrast to a metallic surface, for which it does not matter if it consists of one or multiple plates (see Section~\ref{subsec:multiple-plates}), an IRS must consist of many small elements to reconfigure the local phases with high resolution and thereby achieves a main beam with the desired angle $\theta_r$. This has no negative impact on the maximum magnitude of the scattered field and the beamwidth will not be narrower than in the ideal case of a passive metallic surface. As discussed in Section~\ref{eq:preliminaries}, the incoming wave's
E-field in \eqref{incident:field} induces an electric surface current
in the direction of $\be_x$. This current is adjusted in the IRS by
tuning the surface impedance in each element to obtain a surface phase
profile that approximates that required by the generalized Snell's
law. This operation results in a scattered wave with maximum amplitude
towards $\theta_r$ instead of $\theta_i$.

\begin{lemma}
When using an IRS to reflect a signal in the direction $\theta_r$, the
squared magnitude of the scattered field at an arbitrary observation
angle $\theta_s \in [-\frac{\pi}{2},\frac{\pi}{2}]$ is
	\begin{align} 
	&S_\mathrm{IRS}(r, \theta_s,E_i^2) \nonumber \\
	&=    \pp{\frac{ab}{\lambda}}^{\!\!2} \frac{E_i^2 \cos^2(\theta_i)}{r^2}  
	\pp{ \! \frac{\sin\pp{ \frac{\pi b}{\lambda} (\sin(\theta_s) -\sin(\theta_r)) }}{ \frac{\pi b}{\lambda}(\sin(\theta_s) -\sin(\theta_r))} \! }^{\!2} 
	\end{align} 
at a far-field distance  $r \geq \frac{2 \max\left( a^2,b^2 \right)}{\lambda}$.
\end{lemma}
\begin{IEEEproof}
The negligible thickness of the surface allows us to write the
electric current density on the surface $(z=0, y=y')$ approximately as
$J_x = \frac{2 E_i}{\eta} \cos(\theta_i) e^{-j k \sin (\theta_r) y' }
$ \cite[Eq.~(7-54)]{Balanis2012a} assuming the surface is lossless,
i.e., $E_i^2 \cos(\theta_i) = E_r^2\cos(\theta_r)$. Then, we use the
same steps as in Lemma~\ref{lemma:balanis}.
\end{IEEEproof}
The intercepted power by the IRS is the same as for the perfectly
conducting plate in Section~\ref{eq:preliminaries}, but the maximum of
$S_\mathrm{IRS}(r, \theta_s,E_i^2) $ is achieved at $\theta_s =
\theta_r$ instead of $\theta_s = \theta_i$.

Suppose the transmit power is $P_t$ and the transmitter has antenna
gain $G_t$. Then the relation between  $E_i$ and $P_t$ is
\begin{equation} 
\frac{E^2_i }{2 \eta}= \frac{P_t G_t}{4 \pi d^2_i}, 
\end{equation}
where $E_i$ has unit $\mathrm{Volt/m}$ and $\eta \approx 377 \,
\mathrm{Ohm}$.  Furthermore, assume that the effective area of the
receiver antenna is $ \frac{\lambda^2}{4\pi}G_r$, where $G_r$ is the
antenna gain. Then, the received signal power $P_r$ for a receiver at
far-field distance $r$ in direction $\theta_s$ is
\begin{equation}
P_r(P_t,d_i, r,\theta_s) = \frac{1}{2 \eta} S_\mathrm{IRS} \left(r, \theta_s, \frac{P_t G_t \eta }{2 \pi d^2_i} \right) \left( \frac{\lambda^2}{4\pi}G_r \right).
\end{equation}

\begin{corollary}
	When using an IRS to reflect a signal in the direction $\theta_r$, the pathloss at the far-field distance $r$ is
	\begin{align} \label{eq:general-pathloss}
	&\beta_\mathrm{IRS}(r,d_i,\theta_s) = \frac{P_r(P_t,d_i, r,\theta_s) }{P_t} \nonumber\\
	& = \frac{G_t G_r}{(4\pi)^2 } \pp{\frac{ab}{d_i r}}^2  \cos^2(\theta_i)
	\pp{ \frac{\sin\pp{ \frac{\pi b}{\lambda} (\sin(\theta_s) -\sin(\theta_r)) }}{ \frac{\pi b}{\lambda}(\sin(\theta_s) -\sin(\theta_r))}  }^2
	\end{align}
  and in the ideal case when the receiver has $\theta_s = \theta_r$, the pathloss expression simplifies to 
	\begin{align} \label{eq:ideal-pathloss}
	\beta_\mathrm{IRS}(r,d_i,\theta_r) = \frac{G_t G_r}{(4\pi)^2 } \pp{\frac{ab}{d_i r}}^2  \cos^2(\theta_i).
	\end{align}
\end{corollary}

\begin{remark}
The derivations have assumed the incoming wave and the desired
scattered wave have E-fields that are both parallel to $\be_x$, but
the analysis can be generalized. The end result will still become the
pathloss expression $\beta_\mathrm{IRS}(r,d_i,\theta_r) $ that only
depends on the total effective area $ab \cos(\theta_i)$ of the IRS as
seen from the transmitter.
\end{remark}

\begin{figure}[t!]
	\centering  \vspace{-4mm}
	\begin{overpic}[width=1.05\columnwidth,tics=10,height=2.35in]{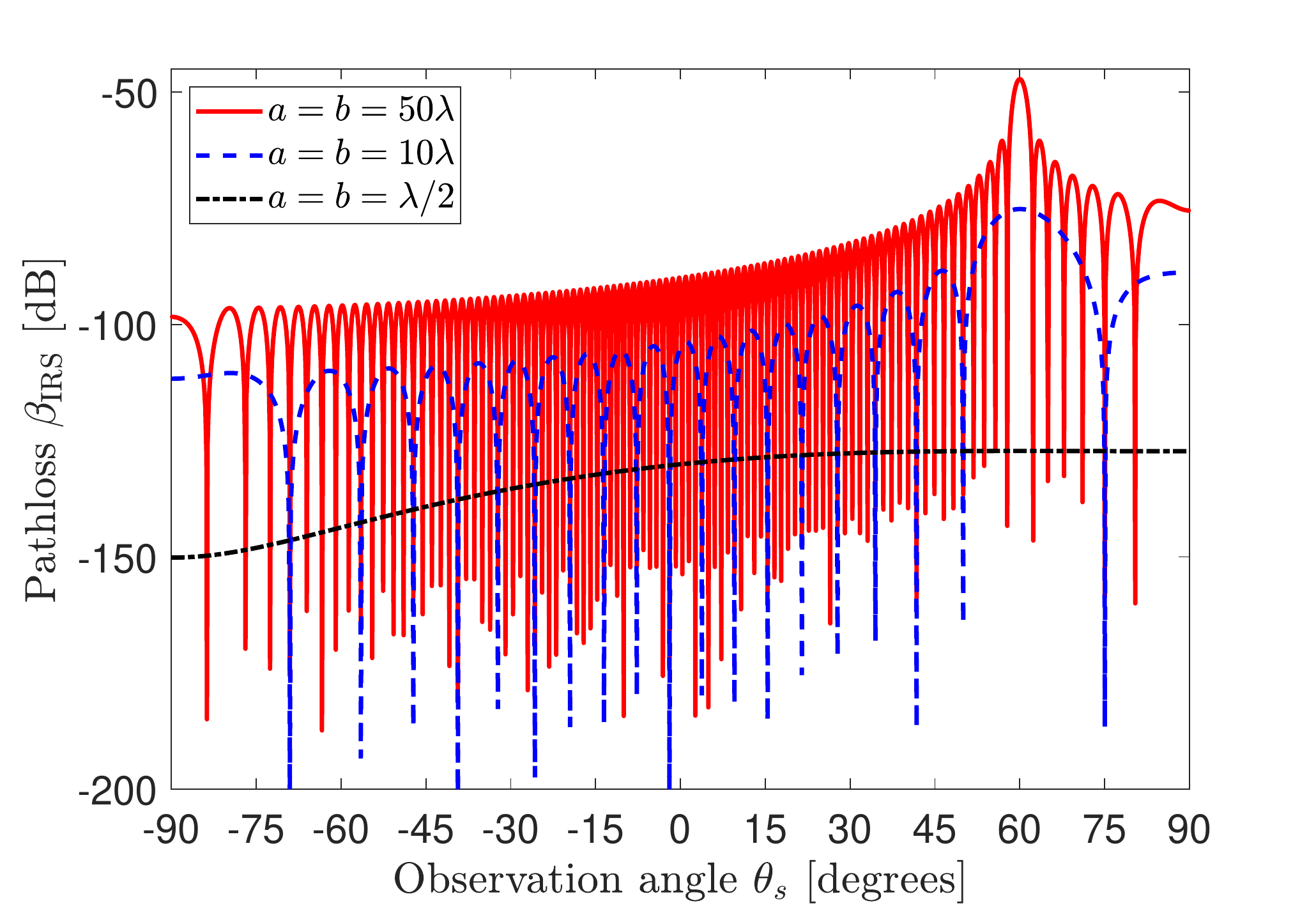}
	\end{overpic} 
	\caption{The pathloss of the reflected path. The angles are $\theta_i=30^\circ$ and $\theta_r = 60^\circ$ where the antenna gains are $G_t =G_r= 5$ dB, the distances are $d_i=50$ and $r =25$ meters.}
	\label{figurePathloss}
	\vspace{-5mm}
\end{figure}

Fig.~\ref{figurePathloss} shows the pathloss
$\beta_\mathrm{IRS}(r,d_i,\theta_s)$ as a function of $\theta_s$ for
different sizes of the IRS and $\theta_r=60^\circ$. The maximum is
achieved at $\theta_s = \theta_r$ and the main beamwidth gets narrower
as the IRS's surface area increases. When the dimension is
sub-wavelength $(\leq \lambda/2)$, the IRS almost acts as a diffuse
scatterer.
\vspace{-5mm}
\subsection{Interpreting an IRS as an Array of Diffuse Scatterers}

As noted above, an IRS of size $a \times b$ consists of many
sub-$\lambda$-sized surface elements. Suppose the IRS consists of $N_a
\times N_b$ elements, each having the size $\frac{a}{N_a} \times
\frac{b}{N_b}$, where $\frac{a}{N_a} , \frac{b}{N_b} \leq \lambda$.
The pathloss between the transmitter and receiver through the
$n^\mathrm{th}$ surface element (assuming the others are removed) is
\begin{align} \label{eq:pathloss-scatter1}
\beta^s_{\mathrm{IRS}}(r,d_i,\theta_r) = \frac{G_t G_r}{(4\pi)^2 } \pp{\frac{ab}{N_a N_b d_i r}}^2  \cos^2(\theta_i) 
\end{align}
since the last term in \eqref{eq:general-pathloss} is approximately
unity for an element of this size (as shown in
Fig.~\ref{figurePathloss}). Note that
$\beta^s_{\mathrm{IRS}}(r,d_i,\theta_r) $ is the same for all $n$
since we have assumed $r \geq \frac{2 \max\left( a^2,b^2
  \right)}{\lambda}$.

Let $\phi_n$ denote the local surface phase of the $n^\mathrm{th}$
element. If it is selected to achieve constructive interference from
all $N=N_a N_b$ surface elements at the receiver, the pathloss between
the transmitter and receiver through the whole IRS is
\begin{equation} \label{eq:phase-aligned-scatterers}
 \left( N \sqrt{\beta^s_{\mathrm{IRS}}(r,d_i,\theta_r)} \right)^2 = \beta_\mathrm{IRS}(r,d_i,\theta_s).
\end{equation}
Hence, we can interpret an IRS as an array of diffuse
scatterers  (each sub-$\lambda$-sized) that phase-align their reflected signals at the receiver and thereby achieve ``anomalous'' reflection.

\vspace{-2mm}

\subsection{System Model for IRS-Supported Communications}

To exemplify how one can derive a physically correct system model for
IRS-supported communication, we consider an LoS setup where
$\sqrt{\beta_\mathrm{sd}} e^{j \phi_\mathrm{sd}}$ is the direct
channel between the single-antenna source and destination. When
including the reflected path from the IRS, we get the received signal
\begin{equation} \label{eq:received-signal1}
	y = \left( \sqrt{\beta^s_{\mathrm{IRS}}} \mathbf{h}^{\Ttran}_\mathrm{sr} \boldsymbol{\Phi} \mathbf{h}_\mathrm{rd} + \sqrt{\beta_\mathrm{sd}} e^{j \phi_\mathrm{sd}} \right) x + w,
\end{equation}
where $\mathbf{h}_\mathrm{sr} = [e^{j\psi^\mathrm{sr}_1},\dots,
  e^{j\psi^\mathrm{sr}_n}, \dots ,e^{j\psi^\mathrm{sr}_N}]^{\Ttran} $
and $\mathbf{h}_\mathrm{rd} = [e^{j\psi^\mathrm{rd}_1},\dots,
  e^{j\psi^\mathrm{rd}_n}, \dots ,e^{j\psi^\mathrm{rd}_N}]^{\Ttran}$
are the normalized LoS channels between the source and IRS and the IRS
and receiver, respectively. The signal $x$ has power $P_t$, $w\sim
\CN(0,\sigma^2)$ is additive noise, and the surface phases of each
surface element are stacked in $\boldsymbol{\Phi} =
\mathrm{diag}\left( e^{j \phi_1}, \dots, e^{j \phi_n}, \dots, e^{j
  \phi_N}\right) $, which is a diagonal matrix. An equivalent way to write \eqref{eq:received-signal1} is
\begin{equation} \label{eq:received-signal2}
	y = \sqrt{\beta^s_\mathrm{IRS}} \sum_{n=1}^{N} e^{j(\psi^\mathrm{sr}_n+ \psi^\mathrm{rd}_n +\phi_n) }
	x + \sqrt{\beta_\mathrm{sd}} e^{j \phi_\mathrm{sd}} x + w.
\end{equation}
The IRS can select $\boldsymbol{\Phi}$ to maximize the received signal
power  \cite{Wu2018a}. If we select $\phi_n \!=\! \phi_\mathrm{sd} -
\psi^\mathrm{sr}_n- \psi^\mathrm{rd}_n$ to phase-align all the signal terms in
\eqref{eq:received-signal2}, we obtain $y = ( N
\sqrt{\beta^s_\mathrm{IRS}} + \sqrt{\beta_\mathrm{sd}} ) e^{j
  \phi_\mathrm{sd}} x + w$ and the signal-to-noise ratio (SNR) is \vspace{-2mm}
\begin{align} 
\mathrm{SNR} = \frac{\left( N \sqrt{\beta^s_\mathrm{IRS}}  + \sqrt{\beta_\mathrm{sd}}\right)^2 P_t}{\sigma^2}  = \frac{\left( \sqrt{\beta_\mathrm{IRS}}  + \sqrt{\beta_\mathrm{sd}}\right)^2 P_t}{\sigma^2},
\end{align}
where the second equality follows from
\eqref{eq:phase-aligned-scatterers}. Hence, the considered
phase-shifts coincide with those achieved by the discretized
generalized Snell's law, with the extra condition that not only the
$N$ terms in the sum in \eqref{eq:received-signal2} are phase-aligned,
but that they are also phase-aligned with the LoS path.

If an IRS is used for ``anomalous'' reflection, there is no need to
take the detour via \eqref{eq:received-signal1}, but we use the IRS's
total pathloss in \eqref{eq:ideal-pathloss} and directly write the
received signal as $y = \sqrt{\beta_\mathrm{IRS}} e^{j
  \phi_\mathrm{IRS}}x + \sqrt{\beta_\mathrm{sd}} e^{j
  \phi_\mathrm{sd}} x + w$, where the phases of the IRS elements have
already been aligned. It then only remains to select the common phase
$\phi_\mathrm{IRS}$ of all IRS elements to equal $\phi_\mathrm{sd}$.
However, there are other scenarios (e.g., with multiple antennas)
where one can start from a system model as
\eqref{eq:received-signal1}.

\vspace{-2mm}

\section{Summary and Relation to Prior Work}

We have used physical optics techniques to derive the pathloss
expression in \eqref{eq:ideal-pathloss} for an IRS that is configured
to reflect an incoming wave from a far-field source towards a receiver in the far-field. Even if the incoming
signal is a plane wave, the reflected signal is a beam with beamwidth
inversely proportional to the size of the IRS. Importantly, the
received signal power is proportional to the square of the IRS area
and to $1/(d_i r)^2$, where $d_i$ is the distance between the
transmitter and IRS, and $r$ is the distance between the IRS and receiver. This disproves the conjecture in \cite{Basar2019a} that
the received power would be proportional to $1/(d_i+r)^2$. That
conjecture might hold for an infinitely large IRS acting or in the near-field, but provably not in the far-field setup studied herein.\footnote{More specifically, \cite[Sec.~II.D]{Basar2019a} states that ``\emph{a metasurface that is capable of shaping the angle and the phase of the reflected signal has a size of the order of $10 \lambda \times 10 \lambda$. This size allows, in general, a metasurface to be viewed as a specular reflector according to geometrical optics.}'' On the contrary, as shown in Fig.~\ref{figureBeamWidth} herein, the beamwidth of the scattered wave from such a surface is $\approx 10^\circ$, hence it is far from a specular reflector, irrespective of $\lambda$, when the source and destination are in the far-field.} In particular, one
cannot use multiple infinite-sized IRS as in \cite{Basar2019a}.
However, the system models used in
\cite{Huang2019,Wu2018a,Bjornson2019IRS} are essentially correct, if the
pathloss of each element are selected according to
\eqref{eq:pathloss-scatter1}.

A practical IRS consists of $N$ sub-wavelength-sized elements that
scatter the incoming signals with unique phase-shifts to achieve
coherent beamforming in a direction of interest. This is rather
similar to a phased array, except that the signal power comes from
another place. The ideal phase-shifts that create a single beam are given by the generalized Snell's law, but if a superposition of multiple beams should be created, the phase-shifts must be explicitly optimized. 

\vspace{-3mm}

\bibliographystyle{IEEEtran}

\bibliography{IEEEabrv,refs}

\end{document}